\documentclass[aps,prx,twocolumn,superscriptaddress,showpacs,floatfix]{revtex4-2}

\usepackage{amsmath,amssymb,graphicx,bm}
\usepackage[colorlinks=true,linkcolor=blue,citecolor=blue,urlcolor=blue]{hyperref}
\usepackage{tikz}
\usetikzlibrary{positioning}
\usepackage{amsmath}
\usetikzlibrary{arrows.meta, decorations.pathmorphing}

\usepackage{orcidlink}

\begin{document}

\title{Fragmented eigenstate thermalization versus robust integrability in long-range models}

\author{Soumya Kanti Pal\,\orcidlink{0009-0008-7226-356X} }
\affiliation{Department of Theoretical Physics, Tata Institute of Fundamental Research, Homi Bhabha Road, Mumbai 400005, India}
\email{Email: soumya.pal@tifr.res.in}
\author{Lea F Santos\, \orcidlink{0000-0001-9400-2709}}
\affiliation{Department of Physics, University of Connecticut, Storrs, Connecticut 06269, USA}
\email{Email: 
lea.santos@uconn.edu}
\date{\today}

\begin{abstract}
Understanding the stability of integrability in many-body quantum systems is key to controlling dynamics and predicting thermalization. While the breakdown of integrability in short-range interacting systems is well understood, the role of long-range couplings -- ubiquitous and experimentally realizable -- remains unclear. We show that in fully connected models, integrability is either robust or extremely fragile, depending on whether perturbations are non-extensive, extensive one-body, or extensive two-body. In contrast to finite short-range systems, where any of these perturbations can induce chaos at finite strength, in fully connected finite models, chaos is triggered by extensive two-body perturbations and even at infinitesimal strength. Chaos develops within energy bands defined by symmetries, leading to a fragmented realization of the eigenstate thermalization hypothesis and clarifying how microcanonical shells can be constructed in such systems. We also introduce a general symmetry-based framework that explains the stability of integrability.
\end{abstract}

\maketitle

Long-range interacting systems have raised fundamental challenges since their earliest studies, as they violate key assumptions of conventional statistical mechanics. Unlike short-range systems, where additivity and extensivity ensure well-defined thermodynamic limits, long-range interactions induce strong correlations across the entire system, requiring a refined theoretical framework. The conditions for thermodynamic stability in such systems were rigorously established in~\cite{Lieb1972,Lebowitz1969}. Building on this foundation, subsequent studies uncovered other unique features, that include ensemble inequivalence~\cite{Barre2001,Campa2009}, quasi-stationary states~\cite{Lynden-Bell1967,Chavanis2006,Mukamel2005,Gupta2010}, and anomalously slow relaxation~\cite{Antoni1995,Mukamel2005,Gupta2010,Bachelard2013}. 

In the quantum domain, long-range interacting systems have revealed a variety of nontrivial phenomena rooted in their intrinsic non-locality and non-additivity~\cite{Defenu2023}. These include excited-state quantum phase transitions~\cite{Caprio2008,Santos2016,Cejnar2021}, finite-energy phase transition in one dimension~\cite{Schuckert2025}, slow entanglement growth~\cite{Pappalardi2018,Lerose2020,Lerose2019}, violations of Lieb-Robinson bounds~\cite{Hauke2013,Eisert2013,Metivier2014,Halati2025}, cooperative shielding~\cite{Santos2016PRL,Celardo2016}, discrete time crystal phases~\cite{Kozin2019,Pizzi2021}, strong prethermalization~\cite{Schutz2014,Schutz2016,Mori2019,Lerose2019,Defenu2021}, unconventional dynamical phase transitions~\cite{Defenu2018,Zunkovic2018,Syed2021,King2023,Gherardini2024,Solfanelli2025}, and the emergence of robust many-body quantum scars~\cite{Lerose2025}. Motivated by these theoretical predictions and enabled by experimental advances in controlling long-range couplings and achieving long coherence times, particularly in arrays of trapped ions~\cite{Lanyon2011,Britton2012,Jurcevic2014,Richerme2014} and Rydberg atoms~\cite{Saffman2010,Scholl2021}, a growing number of experiments have begun to explore far-from-equilibrium dynamics in long-range interacting systems~\cite{Defenu2024}. They have led to direct observations of light-cone violation~\cite{Jurcevic2014,Richerme2014}, entanglement generation~\cite{Bohnet2016}, dynamical phase transitions~\cite{Jurcevic2017}, and long-lived prethermal states~\cite{Kao2021}.

A central open question in systems with long-range interactions concerns the interplay between collective dynamics, prethermalization, and eventual thermalization. In isolated quantum systems, thermalization is widely understood to emerge from the development of many-body quantum chaos~\cite{Borgonovi2016}, which leads to the spreading of correlations and the effective exploration of a large portion of Hilbert space. The mechanism of thermalization is commonly formalized by the eigenstate thermalization hypothesis (ETH),  which states that expectation values of few-body observables in individual energy eigenstates of chaotic Hamiltonians vary smoothly with energy and agree with thermal predictions, while off-diagonal matrix elements are exponentially small in system size~\cite{Alessio2016}.

In systems with short-range interactions, integrability is typically fragile and the addition of even a single impurity can induce chaos~\cite{Santos2004,Gubin2012,Torres2014PRE,TorresKollmar2015,Santos2020,Brenes2020}. This fragility implies that integrability-breaking perturbations drive the system toward chaos at a strength that decays with system size. Whether this paradigm extends to systems with long-range interactions remains an open question. In~\cite{Russomanno2021}, an arbitrarily small perturbation was shown to induce chaos in the presence of long-range interactions, although in~\cite{Sugimoto2022} a finite perturbation threshold was required for thermalization.

To understand how many-body quantum chaos emerges, or fails to emerge, and whether ETH holds in systems with strong long-range interactions, we establish general results for fully connected models. We show that integrability in these systems is either remarkably robust or extremely sensitive, depending on the structure of the perturbation. While non-extensive and extensive one-body perturbations preserve integrability even at finite strength, extensive two-body perturbations can induce chaos already at infinitesimal strength. 

To explain this dichotomy, we develop a general symmetry-based theoretical framework and show that any Hamiltonian invariant under pairwise permutations of lattice sites exhibits an extensive set of conserved quantities and a highly degenerate banded spectrum. This structure determines the stability of integrability. Perturbations that preserve an extensive subset of these symmetries maintain integrability, whereas those that break them generically induce chaos already at infinitesimal strength. The framework applies broadly to systems with finite local Hilbert space dimension and provides a predictive criterion for when integrability is robust and when it breaks down. 

We illustrate these general results using a prototypical Hamiltonian with power-law interactions, which is experimentally realized in trapped-ion platforms. In the fully connected limit, the model becomes mean-field integrable with a spectrum split into highly degenerate energy bands. Additional examples supporting the generality of our findings are presented in the Supplemental Material (SM)~\cite{noteSUPPL}.

\textit{Model and spectrum.--} We consider the one-dimensional system with open boundary conditions described by the following spin-1/2 Hamiltonian with $L$ sites~\cite{Jurcevic2014},
\begin{align} 
\label{eq:LRTFIM}
    \hat{H} = \mathcal{N}_\alpha \sum_{i>j=1}^L J \frac{\hat{\sigma}_i^x \hat{\sigma}_j^x}{|i-j|^{\alpha}} + h \sum_{i=1}^L \hat{\sigma}_i^z,
\end{align}
where $\hat{\sigma}_i^{x,z}$ are Pauli matrices at site $i$, and $\mathcal{N}_\alpha=1/L^{(1-\alpha})$ is the thermodynamic Kac's scaling that ensures energy extensivity for $\alpha<1$. Throughout the paper with fix $J=1$ and $h=1$. The eigenvalues and eigenstates of $\hat{H}$ are denoted by $E_n$ and $|n \rangle$.

The model is integrable in both limits: for nearest-neighbor interaction ($\alpha \to \infty$), when it becomes the transverse-field Ising model, and for all-to-all couplings ($\alpha = 0$), which has also been experimentally realized~\cite{Li2023}. In this case, Hamiltonian~\eqref{eq:LRTFIM} becomes equivalent to that of the Lipkin-Meshkov-Glick (LMG) model, 
\begin{align}
\label{eq:collective spin Hamiltonian}
    \hat{H}^{\alpha=0} = \frac{2J}{N} \hat{S}_x^2 + 2h \hat{S}_z - 2J, 
\end{align}
where the collective spin operators are defined as $\hat{S}_{x,y,z} = (1/2) \sum_{i=1}^L \hat{\sigma}_i^{x,y,z}$. The LMG Hamiltonian $\hat{H}^{\alpha=0}$ has $SU(2)$ symmetry, as the total spin $\hat{S}^2 = \hat{S}_x^2 + \hat{S}_y^2 + \hat{S}_z^2$ is conserved. 

\begin{figure}[t]
    \centering
    \includegraphics[width=1.0\linewidth]{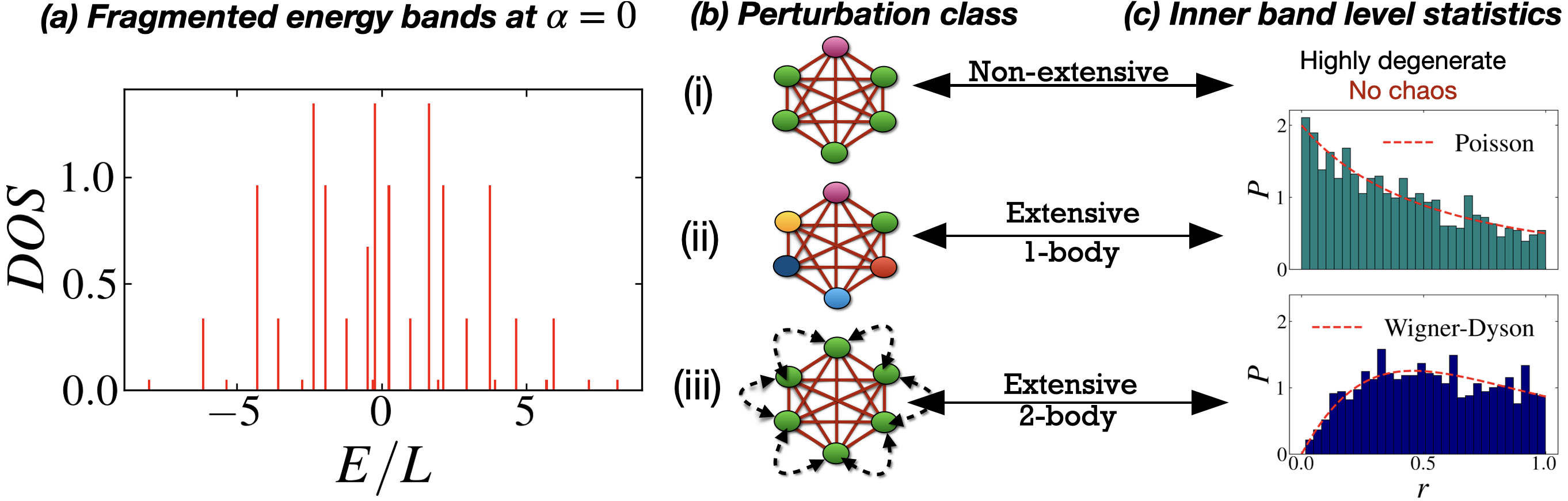}
    \caption{In the all-to-all coupling limit ($\alpha=0$), (a) the density of states (DOS) is split into energy bands when the system ($L=8$) is subjected to (b) perturbations from any of the three categories. However, the (c) level statistics within the most populated energy band reveal that only class (iii) perturbations induce many-body quantum chaos, as indicated by the Wigner-Dyson level spacing distribution for system ($L=14,~J=1,~h=1$). The representative examples from each class are: (i) $\delta \, \sigma_{L/2}^z$, (ii) $\sum_i^L h_i \sigma_{i}^z$ with $h_i \in [-\delta,\delta]$, (iii) $\delta \sum_{i=1}^{L-1} \hat{\sigma}_i^x \hat{\sigma}_{i+1}^x$, where $\delta =10^{-4}$. Parity and inversion symmetries were taken into account accordingly. 
    }
    \label{fig01}
\end{figure}

The spectrum of $\hat{H}^{\alpha=0}$ consists of highly degenerate energy bands characterized by the total spin quantum number $s$ corresponding to the eigenvalue of $\hat{S}^2$.  For even $L$, $s$ ranges from $0$ to $L/2$, resulting in $(L/2+1)^2$ distinct bands (see SM~\cite{noteSUPPL}). The band structure of the density of states (DOS) is  shown in Fig.~\ref{fig01}(a). We use the term ``fragmented'' to refer to this banded structure, where the Hilbert space decomposes into symmetry sectors that are not mixed by the integrable Hamiltonian.

{\em Perturbations and level statistics.--} To systematically assess the stability of integrability in $\hat{H}^{\alpha=0}$, we classify perturbations into three experimentally relevant categories.
The energy-band structure of the DOS in Fig.~\ref{fig01}(a) remains unchanged under an infinitesimal perturbation $\hat{V}$ of strength $\delta$ from any of the three classes. They are schematically depicted in Fig.~\ref{fig01}(b) and categorized as follows: 
(i) Non-extensive perturbations include one-body or two-body terms, local or non-local, whose support does not scale with system size, such as $\hat{\sigma}_{L/2}^z$, $\hat{\sigma}_{L/2}^z\hat{\sigma}_{L/2 +1}^z$, or $\hat{\sigma}_{1}^z\hat{\sigma}_{L}^x$. (ii) Extensive one-body perturbations consist of uniform or disordered fields with support growing with system size, such as $\sum_{i=1}^L\hat{\sigma}_{i}^x$, $\sum_{i=1}^{L/2}\hat{\sigma}_{i}^z$, or $\sum_{i=1}^L h_i \hat{\sigma}_i^z$ where $h_i \in [-\delta, \delta]$ are random numbers. (iii) Extensive two-body perturbations include local two-body terms with system-size-dependent support, such as $ \sum_{i=1}^{L} \hat{\sigma}_i^z \hat{\sigma}_{i+1}^z$, $\sum_{i=1}^{L} h_i \hat{\sigma}_i^x \hat{\sigma}_{i+1}^x$ or infinitesimal changes to the power-law exponent, $\alpha \to \alpha + \delta\alpha$.

Despite sharing equivalent DOS, the three classes of infinitesimal perturbations yield markedly different spectral properties. The presence of correlated eigenvalues, as in random matrix theory~\cite{MehtaBook}, is a widely used diagnostic of quantum chaos in isolated systems. In particular, short-range spectral correlations can be quantified through the analysis of adjacent energy levels, for example via the distribution of the ratio  $r_n = \min{({\cal S}_n,{\cal S}_{n-1})}/\max{({\cal S}_n,{\cal S}_{n-1})}$ of consecutive level spacings, ${\cal S}_n=E_{n+1}-E_n$, \cite{Atas2013}. Quantum chaos is signaled by level repulsion and Wigner-Dyson statistics~\cite{Guhr1998}, whereas integrable systems exhibit uncorrelated levels, with Poissonian spacing distributions, or degeneracies arising from conserved quantities. In Fig.~\ref{fig01}(c), we show the level spacing distributions computed within the most populated energy band for representative perturbations from each class. In general, integrability is preserved under class (i) and class (ii) with either homogeneous perturbations or random perturbations in the transverse direction -- even at finite strength. As seen in Fig.~\ref{fig01}(c), class (i) perturbations barely lift degeneracies, while class (ii) does, despite maintaining integrability, as indicated by the Poisson distribution. In stark contrast, class (iii) perturbations induce Wigner-Dyson statistics even for infinitesimal strength, signaling the onset of quantum chaos.

\begin{figure*}
    \centering
    \includegraphics[width=1.0\linewidth]{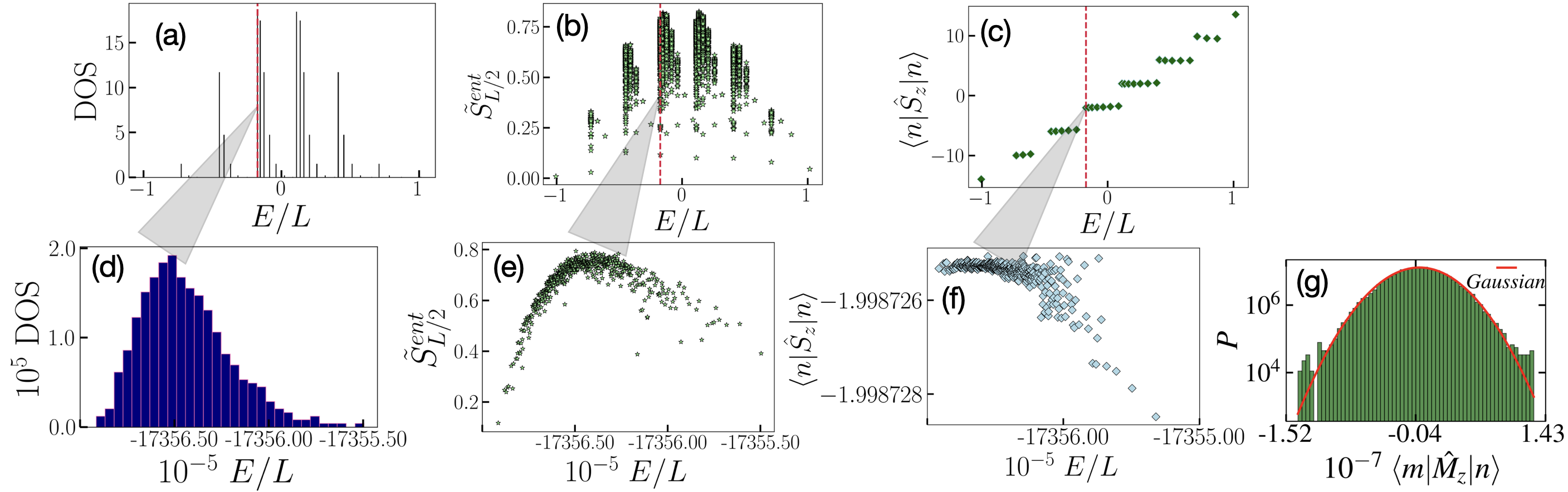}
    \caption{Verification of the eigenstate thermalization hypothesis (ETH) for Hamiltonian~\eqref{eq:LRTFIM} with $\alpha=10^{-4}$, $L=14$. The top panels show results for the full energy spectrum, with the vertical red line marking the most populated band analyzed in the bottom panels. (a) and (d): Density of states, (b) and (e): Entanglement entropy, (c) and (f): Eigenstate expectation values of $\hat{S}_z$, and (g): Off-diagonal elements of $\hat{S}_z$ for 200 eigenstates in the middle of the selected energy band. (f) and (g): Diagonal and off-diagonal ETH, respectively, are satisfied. Parity and inversion symmetries are taken into account. }
    \label{fig02}
\end{figure*}

{\em Chaos and eigenstate thermalization.--} The emergence of chaos in long-range systems as the interaction exponent $\alpha$ is tuned from zero to small non-zero values [class (iii)] was previously reported in Ref.~\cite{Russomanno2021}. Despite the Wigner-Dyson level spacing distribution, which is a hallmark of quantum chaos, that paper along with Ref.~\cite{Sugimoto2022} suggested that the eigenstates may fail to satisfy ETH and microcanonical energy shells could no longer be properly defined~\cite{Russomanno2021}. More recently, this violation was reinforced from an open quantum system perspective in \cite{Mattes2025}, and efforts to restore thermalization for long-range systems have also been proposed~\cite{winter2025}. In this work, we revisit the claim of ETH breakdown in quantum systems with strong long-range interactions and, in contrast, provide evidence that ETH does hold, although in a more subtle, sector-dependent form.

For an infinitesimally small $\alpha$, the band structure of the DOS remains intact [Fig.~\ref{fig02}(a)], although the degeneracies within each band are lifted. In the weak perturbation regime, the bands can still be approximately characterized by the total spin quantum number, 
since the commutator norm goes to zero in the limit $\alpha\to 0$, i.e., $\lim_{\alpha \to 0} ||[\hat{S}^2, \hat{H}^{\alpha}]|| \to 0$. As a result, a proper analysis of ETH must be performed within these individual sectors associated with the approximate conservation of $\hat{S}^2$. Constructing energy shells that combine states across different sectors can obscure the existence of chaotic states and misleadingly indicate a breakdown of ETH. This is indeed suggested by the fragmented results for the scaled half-chain entanglement entropy,
\begin{equation}
  \tilde{S}^{\mathrm{ent}}_{L/2} = - \frac{2}{L\ln 2} \text{Tr}_{L/2}\left[ \hat\rho_{L/2} \ln{(\hat\rho_{L/2})} \right]  
\end{equation}
in Fig.~\ref{fig02}(b), and the eigenstate expectation values, $\langle n| \hat{S}_z |n\rangle$, of the total magnetization in the $z$-direction in Fig.~\ref{fig02}(c), both shown as a function of the energy levels of the entire spectrum. 

This seemingly violation of ETH disappears when the analysis is restricted to individual energy bands. As shown in Fig.~\ref{fig02}(d), the DOS for the most populated band closely resembles a Gaussian distribution, as typical of many-body quantum systems~\cite{Brody1981}. Consistent with the presence of chaotic eigenstates~\cite{ZelevinskyRep1996,Borgonovi2016}, which underlie the validity of ETH~\cite{Santos2010PRE}, the entanglement entropy of the eigenstates within the band, and away from its edges, exhibits a smooth dependence on energy in Fig.~\ref{fig02}(e). This behavior is mirrored by the smooth variation of the eigenstate expectation values of $\hat{S}_z$, as shown in Fig.~\ref{fig02}(f). Additionally, the distribution of the off-diagonal matrix elements of $\hat{S}_z$ for the eigenstates in the middle of the selected energy band, displayed in Fig.~\ref{fig02}(g), is well-approximated by a Gaussian, in agreement with chaotic eigenstates and thus, with the predictions of off-diagonal ETH~\cite{Beugeling2015,LeBlond2019}.

These observations provide strong evidence that ETH is satisfied within individual energy bands. Furthermore, we have confirmed that this conclusion holds robustly for other models and perturbations within class (iii). See, as an example, the case of the all-to-all XXZ Hamiltonian in SM~\cite{noteSUPPL}.

{\em Origin of robust integrability and general framework.--} The robustness of integrability in the fully connected limit ($\alpha=0$) originates from the permutation symmetry of the Hamiltonian. The system is invariant under pairwise permutations $P_{ij} = \frac{1}{2}\left(\mathbb{I}_i \otimes \mathbb{I}_j +  \hat{\vec\sigma}_i \cdot  \hat{\vec\sigma}_j\right) $ between any two sites $i$ and $j$, which generate an extensive set of conserved quantities and lead to a highly degenerate banded energy spectrum. 

The effect of perturbations can be understood in terms of symmetry reduction. An intuitive picture is to view the system as a fully connected graph with $L$ equivalent nodes; perturbations act as ``impurities'' that distinguish subsets of nodes and reduce permutation symmetry~\cite{noteSUPPL}.

Class (i): A perturbation acts on $m\ll L$ sites, breaking the full permutation symmetry, but preserving permutations among the remaining $L-m$ sites, leaving $\binom{L-m}{2}$ independent conserved charges. The number of conserved quantities, therefore, remains extensive, and the degeneracy structure is only partially lifted, ensuring the persistence of integrability. 

Class (ii): Extensive one-body perturbations partition the system into subsets that retain independent permutation symmetries, thereby preserving an extensive number of conserved quantities. This is straightforward for homogeneous fields, and remains valid for transverse inhomogeneous fields, $\sum_{i=1}^L h_i \hat{\sigma}^z_i$, where the model is likely related to integrable Richardson-Gaudin models~\cite{noteSUPPL}. In contrast, for longitudinal inhomogeneous fields, $\sum_{i=1}^L h_i \hat{\sigma}^x_i$, degeneracies within the energy bands are lifted more effectively, leading to a crossover toward Wigner-Dyson statistics as the system size increases
~\cite{noteSUPPL}.

This mechanism of robust integrability is not specific to the model in Eq.~(\ref{eq:LRTFIM}). More generally, any Hamiltonian invariant under pairwise permutations,
\begin{align} \label{eq:general_symmetry}
    P_{ij} \hat{H} P_{ij} = \hat{H}, \quad \forall\, i,j \in \{1,\dots,L\},
\end{align}
where $P_{ij}$ exchanges the local Hilbert spaces $\mathbb{C}^d$ at sites $i$ and $j$, exhibits an extensive set of conserved quantities and banded spectrum, independently of the local dimension $d$. In this case, perturbation from class~(iii) can give rise to chaos inside the energy bands. For instance, the parent integrable Hamiltonian in~\cite{Abdelshafy2025} belongs to this symmetry class, and chaos is induced there by perturbations of class~(iii).

{\em Onset of chaos from perturbation theory.--} To identify the mechanism by which integrability breaks down near the fully connected limit, we use degenerate perturbation theory around $\alpha=0$. For a perturbation $\hat V$ from one of the three classes introduced above, the Hamiltonian reads
\begin{align}
    \hat H_{\rm pert}=\hat H^{\alpha=0}+\hat V .
    \label{eq:Hpert}
\end{align}
We focus on the most populated degenerate energy band of the unperturbed Hamiltonian $\hat{H}^{\alpha=0}$, with eigenstates $|n_0\rangle$, and diagonalize the projected perturbation matrix
$\langle m_0|\hat V|n_0\rangle $.
Its eigenvalues $\lambda_{n_0}$ give the first-order energy corrections within the band.  If the eigenvalues $\lambda_{n_0}$ are all zero or show internal degeneracies, the original degeneracy of the band is preserved or partially lifted, suggesting that integrability is maintained. In contrast, a fully non-degenerate $\lambda_{n_0}$ accompanied by level repulsion signals the breakdown of integrability and the onset of quantum chaos. Degenerate perturbation theory thus provides a diagnostic of the interplay between symmetry, degeneracy lifting, and the onset of chaos. 

For class (i) and class (ii), this perturbative analysis is consistent with the symmetry-based picture discussed above. The projected perturbation matrix either vanishes or retains a structured degeneracy, in agreement with the persistence of integrability (see SM~\cite{noteSUPPL}).

Class (iii): Extensive two-body perturbations break the permutation symmetry responsible for the degenerate bands and induce couplings between the states within each band. As a result, the perturbation matrix lifts the degeneracy already at first order and its eigenvalues exhibit Wigner-Dyson statistics~\cite{noteSUPPL}. This happens for tiny perturbation strengths, showing that level repulsion emerges before different unperturbed bands hybridize. Chaos sets in within each band, which explains why extensive two-body perturbations destabilize the $\alpha=0$ integrable point at infinitesimal strength. This perturbative mechanism also clarifies the validity of eigenstate thermalization within individual energy bands.


{\em Conclusion.--} We have shown that integrability in a fully connected many-body quantum system is governed by a general symmetry-based mechanism. Hamiltonians invariant under pairwise permutations exhibit an extensive set of conserved quantities and a fragmented, highly degenerate spectrum, which renders integrability robust against broad classes of perturbations. In contrast, perturbations that break these symmetries extensively lift the degeneracies within each band and induce quantum chaos already at infinitesimal strength.

Drawing a heuristic analogy with classical chaos, where localized chaotic regions in phase space emerge and progressively expand until they merge together~\cite{chirikov2008}, quantum chaos in fully connected systems unfolds in a similarly fragmented fashion. It first emerges within individual energy bands associated with approximate conserved quantities, and as the strength of the integrability-breaking perturbation increases, these chaotic regions broaden and eventually coalesce, signaling a global breakdown of integrability. 

The emergence of 
ETH satisfied within symmetry-defined sectors in the strong long-range regime provides a natural framework for defining microcanonical energy shells. 
Building on this perspective, an important theoretical and experimental extension of this work is to investigate the nonequilibrium dynamics and thermalization timescales near $\alpha = 0$, comparing the behavior across different perturbation classes. In particular, it is important to contrast the evolution of initial states confined to a single energy band with that of states that span multiple bands.

{\it Acknowledgements.--} The authors thank Shamik Gupta for inspiring discussions on long-range systems. S.~K.~P. acknowledges helpful discussions with Marcello Dalmonte, Tobias Schatez, Vyshakh B. R., and Shreya Vardhan. L.~F.~S. thanks start-up funding from the University of Connecticut. S.~K~.P. is supported at the Tata Institute of Fundamental Research (TIFR) through a graduate fellowship from the Department of Atomic Energy (DAE), India. The authors gratefully acknowledge the ``School on Classical and Quantum Long-range Interacting Systems 2024" held at TIFR Mumbai, where this project was initiated.

\bibliography{ms}



\onecolumngrid

\vspace*{0.5cm}

\begin{center}

{\large \bf Supplemental Material: 
\\Fragmented eigenstate thermalization versus robust integrability in long-range models}\\

\vspace{0.6cm}

Soumya Kanti Pal$^1$, Lea F. Santos$^2$\\

$^1${\it Department of Theoretical Physics, Tata Institute of Fundamental Research, Homi Bhabha Road, Mumbai 400005, India}

$^2${\it Department of Physics, University of Connecticut, Storrs, Connecticut 06269, USA}

\end{center}

\vspace{0.6cm}


\setcounter{section}{0}
\renewcommand\thesection{\Roman{section}}
\renewcommand\thesubsection{\thesection.\Alph{subsection}}
\renewcommand{\thefigure}{S\arabic{figure}}
\renewcommand{\theequation}{S\arabic{equation}}

This Supplemental Material provides additional figures, analytical arguments, and numerical evidence that support and extend the findings of the main text. It is organized into the following parts. In Sec.~I, we show that introducing a small interaction exponent $\alpha$ can be interpreted as a nonlocal perturbation to the fully connected Hamiltonian at $\alpha = 0$, and we quantify its perturbative strength. In Sec.~II, we describe the banded structure of the density of states at $\alpha = 0$, arising from total spin conservation, and characterize the degeneracies and scaling of the most populated symmetry sector. In Sec.~III, we develop a degenerate perturbation theory framework within a single band to explain how different classes of perturbations---non-extensive, extensive one-body, and extensive two-body---either preserve integrability or induce chaos, supported by level statistics, spectral form factors, and ETH indicators. In Sec.~IV, we provide a symmetry-based interpretation of the band structure and its robustness using Schur--Weyl duality and a graphical picture of permutation symmetry breaking. Finally, in Sec.~V, we present an additional example based on a long-range XXZ model, highlighting the emergence of fragmented ETH and clarifying the role of disorder through its connection to Richardson--Gaudin integrability. 

\section{Why tuning $\alpha$ is a non-local perturbation to $\hat{H}^{\alpha=0}$}
\label{Supp1}

\begin{figure}[b]
    \centering
    \includegraphics[width=0.5
    \linewidth]{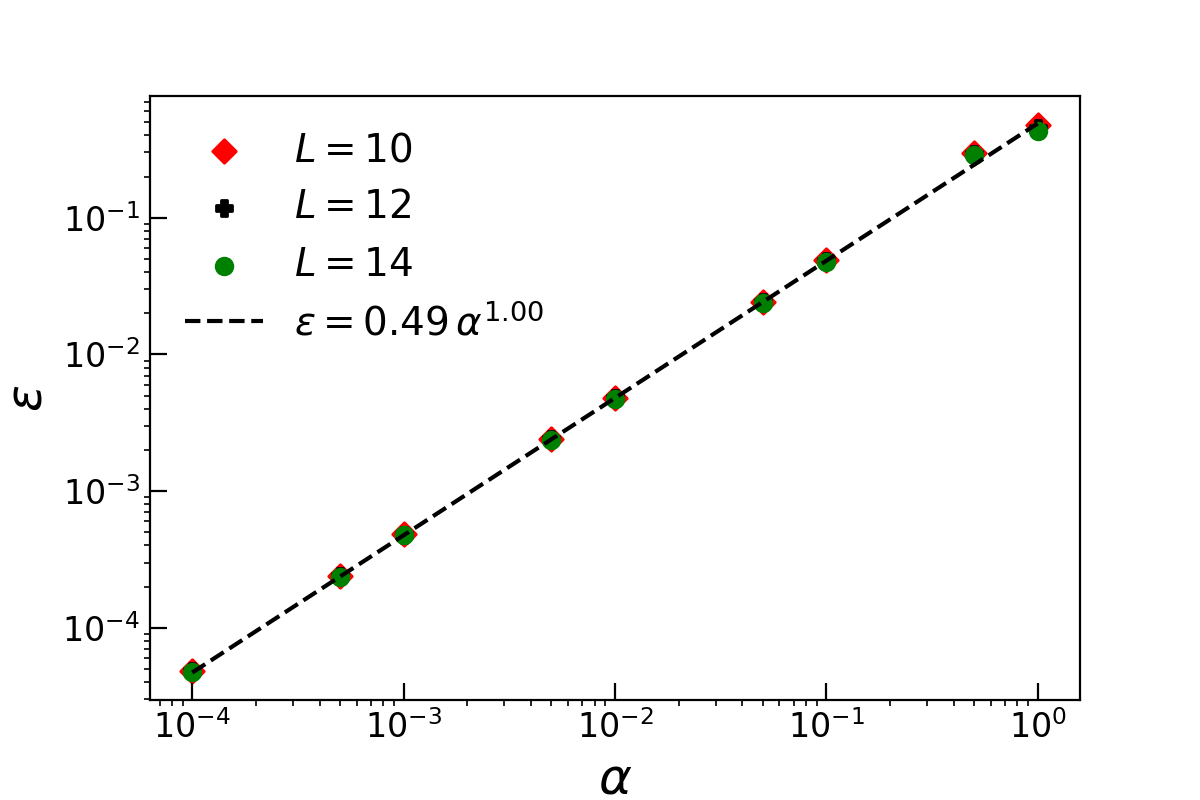}
    \caption{For the Hamiltonian given in Eq.~(1) of the main text with $\alpha=0$, this figure demonstrates that introducing a small but finite $\alpha$ acts as a perturbation, since the ratio of the norms can be fitted to the form $\varepsilon = 0.49\, \alpha^{1.0}$.
}
    \label{fig:ratio-norm}
\end{figure}

Starting from the Hamiltonian in Eq.~(1) of the main text, we consider a small deviation $\alpha \ll 1$ from $\alpha = 0$. The Hamiltonian can then be expressed as
\begin{align}
    \hat{H}^{\alpha\ll 1} = \hat{H}^{\alpha=0} + \underbrace{(\hat{H}^{\alpha\ll 1} - \hat{H}^{\alpha=0})}_{\equiv\,\hat{V}_\alpha},
\end{align}
where the second term, $\hat{V}_\alpha$, acts as a non-local perturbation. To quantify the perturbative nature of $\hat{V}_\alpha$ for $\alpha<1$, we define
\begin{align} \label{eq:ratio-norm}
    \varepsilon(\alpha) = \frac{|| \hat{V}_\alpha ||}{|| \hat{H}^{\alpha=0} ||},
\end{align}
where the Hilbert-Schmidt norm is given by $||A|| = \sqrt{\sum_i \lambda_i^2}$, with $\lambda_i$ denoting the eigenvalues of the Hermitian matrix $A$. As shown in Fig.~\ref{fig:ratio-norm}, we find $\varepsilon(\alpha) < 1$ for $\alpha<1$, validating the interpretation of $\hat{V}_\alpha$ as a perturbation to $\hat{H}^{\alpha=0}$.

\section{Total spin conservation at $\alpha=0$:  Density of states Band structures}

The Hamiltonian in Eq.~(2) of the main text, corresponding to the $\alpha = 0$ limit of the Hamiltonian in Eq.~(1) of the main text, presents $SU(2)$ symmetry, which implies conservation of the total spin, $[\hat{S}^2, \hat{H}^{\alpha=0}] = 0$, with $\hat{S}^2 \equiv \hat{S}_x^2 + \hat{S}_y^2 + \hat{S}_z^2$. This symmetry leads to a highly degenerate spectrum, with the density of states exhibiting distinct energy bands, each containing a large number of degenerate eigenstates. 

Since $\hat{S}^2$ is a symmetry of the Hamiltonian, one can characterize each energy band (namely the eigenstates in the band) by its total spin value $s$. 
For a given $L$ (taken to be even), the total spin can take values $s=0,1,\ldots, L/2$. Due to the presence of the transverse field, each $s$-sector further splits into $(2s+1)$ bands. 
Therefore, for a given $L$, the total number of bands, each corresponding to a distinct energy eigenvalue, is given by
\begin{equation}
\sum_{s=0}^{L/2} (2s+1) = \left( \frac{L}{2} + 1 \right)^2.
\end{equation}

The number of eigenstates in each energy band depends only on the corresponding total spin quantum number $s$. This number corresponds to the degeneracy associated with total spin resulting from the addition of angular momenta of $L$ spin-$\tfrac{1}{2}$ particles. Following the standard rules encoded in Catalan's triangle for combining spin-$\tfrac{1}{2}$ particles, we obtain that for any $s < L/2$, the number of eigenstates in each of the $(2s + 1)$ bands is given by
\begin{align}
    n(s,L) = \frac{2s + 1}{L + 1} \binom{L + 1}{\frac{L}{2} - s},
\end{align}
whereas for $s = L/2$, corresponding to the fully symmetric sector, there is no degeneracy, so each of the $(L+1)$ eigenstates is non-degenerate. 

\begin{figure}[h]
    \centering
    \includegraphics[width=0.9\linewidth]{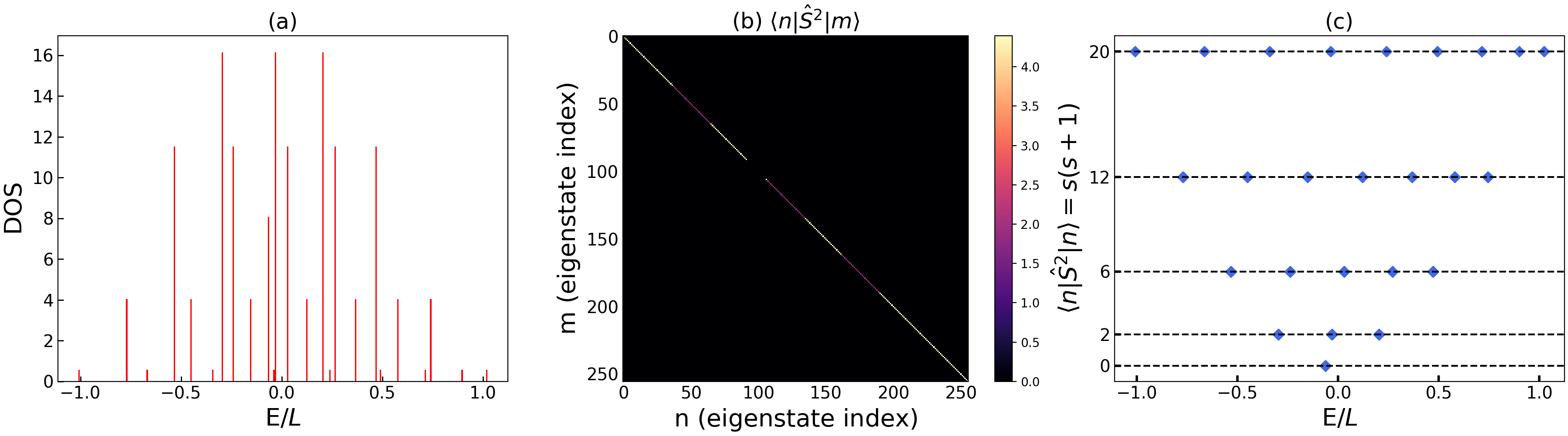}
    \caption{For the fully connected limit of the long-range transverse field Ising model with the Hamiltonian given in Eq.~(2) of the main text and  $L = 8$ spins, panel~(a) shows the density of states (DOS), revealing 25 distinct energy bands. Panel~(b) confirms that $\hat{S}^2$ is diagonal in the eigenstates of the Hamiltonian in Eq.~(2) of the main text. Panel~(c) shows that each band is uniquely identified by its corresponding total spin quantum number $s$.}
    \label{fig:band-alpha-analysis}
\end{figure}

In Fig.~\ref{fig:band-alpha-analysis}, we illustrate the band structure of the density of states (DOS) for $L = 8$, along with the corresponding characterization of each band by the total spin quantum number $s$. 

\begin{figure}
    \centering
    \includegraphics[width=0.35\linewidth]{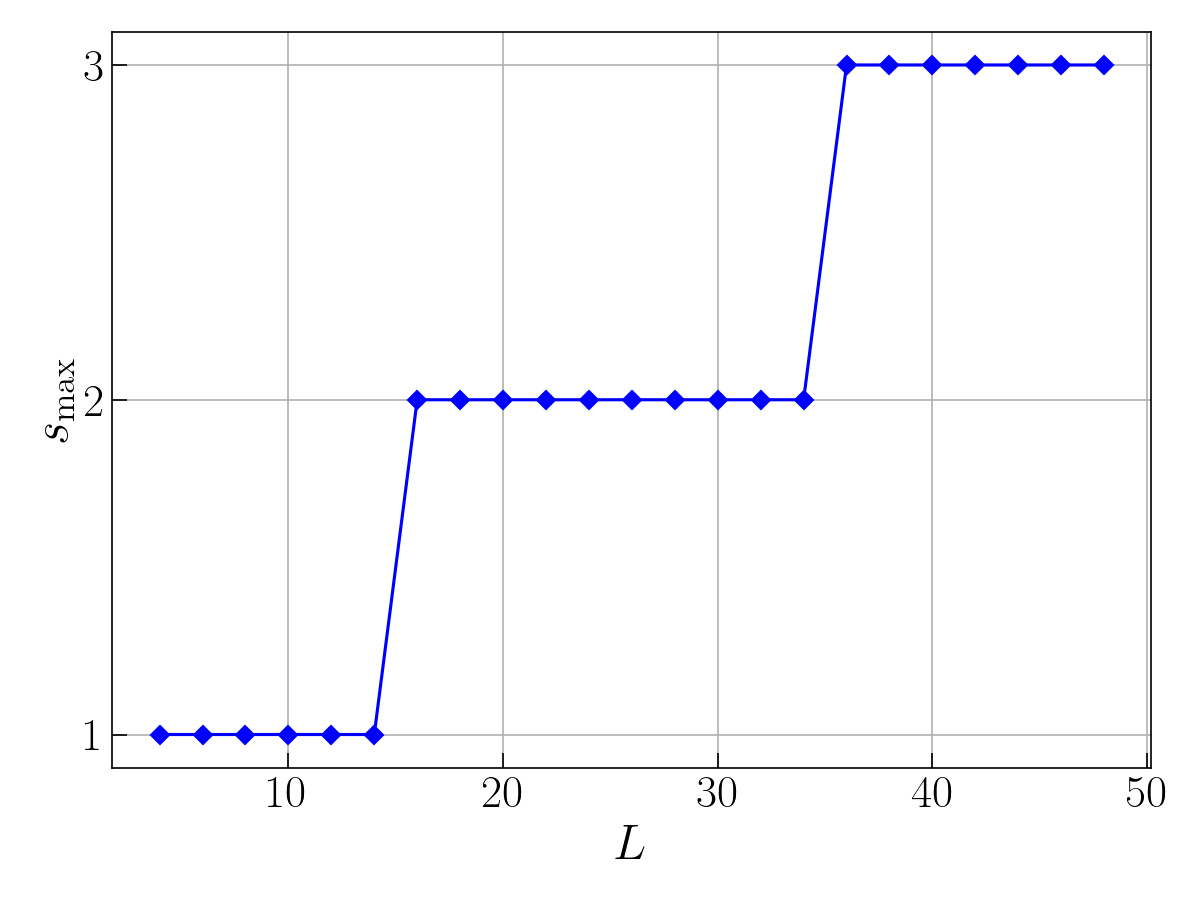}
    \caption{For the Hamiltonian at $\alpha=0$ in Eq.~(2) of the main text, we present the  maximally populated $s_\mathrm{max}$-sector for increasing $L$. Even for numerically accessible system sizes, we find variations of $s_\mathrm{max}$ with $L$.  }
    \label{fig:smax}
\end{figure}

An additional key point is to identify, for a given system size $L$, the sector $s$ that contains the largest number of states. Determining this $s_{\mathrm{max}}$ is important, because studies of quantum chaos and ETH typically focus on the most populated symmetry sector. This tasks amounts to maximizing $n(s,L)$ with respect to $s$, which can be straightforwardly addressed numerically. As shown in Fig.~\ref{fig:smax}, the size of the most populated sector, $s_{\mathrm{max}} \equiv \max_{s} n(s,L)$, varies as a function of $L$.

\section{Degenerate perturbation theory: Why a certain class gives chaos while others retain integrability? }

In this section, we provide a first-principles explanation for the emergence of either integrability or chaos in the cases of (i) non-extensive, (ii) extensive one-body, and (iii) extensive two-body perturbations. To this end, we employ degenerate perturbation theory with the setup 
\begin{align}
    \hat{H} = \hat{H}^{\alpha=0} + \hat{V},
\end{align}
where $\hat{H}^{\alpha=0}$ is the unperturbed Hamiltonian as given in Eq.~(2) of the main text, and $\hat{V}$ denotes the perturbation belonging to one of the three classes above. 
The method for analyzing the effect of the perturbations is as follows:
\begin{itemize}
    \item First, we pin down the symmetries of the Hamiltonian $\hat{H}$ (perturbation included) and  numerically obtain the spectrum of the unperturbed Hamiltonian $\hat{H}^{\alpha=0}$ in a symmetry-resolved manner.  

    \item Next, we numerically identify the most populated energy band in the DOS of $\hat{H}^{\alpha=0}$, with energy $E_{n_0}$, and restrict our analysis to the corresponding eigenstates $\{ | n_0 \rangle \}$ within this band.

    \item For each pair of states $|m_0\rangle$, $|n_0\rangle$ in this band, we compute the matrix elements of the perturbation, $\langle m_0 | \hat{V} | n_0 \rangle$.

    \item We diagonalize this perturbation matrix. The resulting eigenvalues $\{ \lambda_{n_0}^i \}$ yield the first-order energy corrections as $E_{n}^i = E_{n_0} + \lambda_{n_0}^i$.
\end{itemize}

If most of the eigenvalues $\lambda^i_{n_0}$ vanish, the degeneracy is largely preserved, indicating integrability. If the eigenvalues are non-zero but exhibit degeneracies or clustering, the original band splits into multiple subbands, and the system may retain integrable features. On the other hand, if the degeneracy is lifted, the analysis of level statistics determines whether the system is chaotic (level repulsion) or not. In the next subsections, we consider the three classes separately. 

\subsection{Class (i): Non-extensive}
For this class, we consider perturbations---either one-body or two-body---whose spatial support remains fixed as the system size increases.\\

{\bf One-body perturbations.} We introduce a single-site impurity at a bulk site, either in the longitudinal or transverse direction. For a longitudinal impurity, the perturbation takes the form $ \delta \hat{\sigma}^x_{L/2} $. In this case, we find that all first-order corrections vanish, i.e., $ \lambda_{n_0}^i = 0 $ for all $ i $, indicating that the degeneracy of the unperturbed band remains intact. In contrast, a transverse impurity, given by $ \delta \hat{\sigma}^z_{L/2} $, 
leads to nonzero values of $ \lambda^i_{n_0} $. However, the spectrum of the perturbation matrix $ \langle m_0 | \hat{V} | n_0 \rangle $ exhibits internal degeneracies, leading to a splitting of the original degenerate band into several subbands, each with its own residual degeneracy. 

Although perturbation theory is strictly valid only for $ \delta \ll 1 $, we observe that this qualitative structure persists even for larger values of $ \delta $. These results explain the persistence of integrability in the presence of such local perturbations. We also note that in the other integrable limit $ \alpha \to \infty $, a longitudinal impurity of the same kind would instead induce chaos, in stark contrast to the behavior observed here.\\

{\bf Two-body perturbations.} In the bulk, we consider a single two-body interaction term, either in the longitudinal or transverse direction, given respectively by $ \hat{\sigma}^x_{L/2-1} \hat{\sigma}^x_{L/2} $ or $ \hat{\sigma}^z_{L/2-1} \hat{\sigma}^z_{L/2} $. Similar to the one-body case, the eigenvalues $ \{ \lambda_{n_0}^i \} $ of the perturbation matrix $ \langle m_0 | \hat{V} | n_0 \rangle $ exhibit degeneracies, indicating that a single band splits into multiple subbands, each retaining some degree of degeneracy. This spectral structure reflects the persistence of integrability despite the perturbation. We also note that the choice of the specific pair $(L/2-1, L/2)$ is not essential; any arbitrary pair in the bulk produces qualitatively similar results.

\begin{figure}[t]
    \centering
   \includegraphics[width=1.0\linewidth]{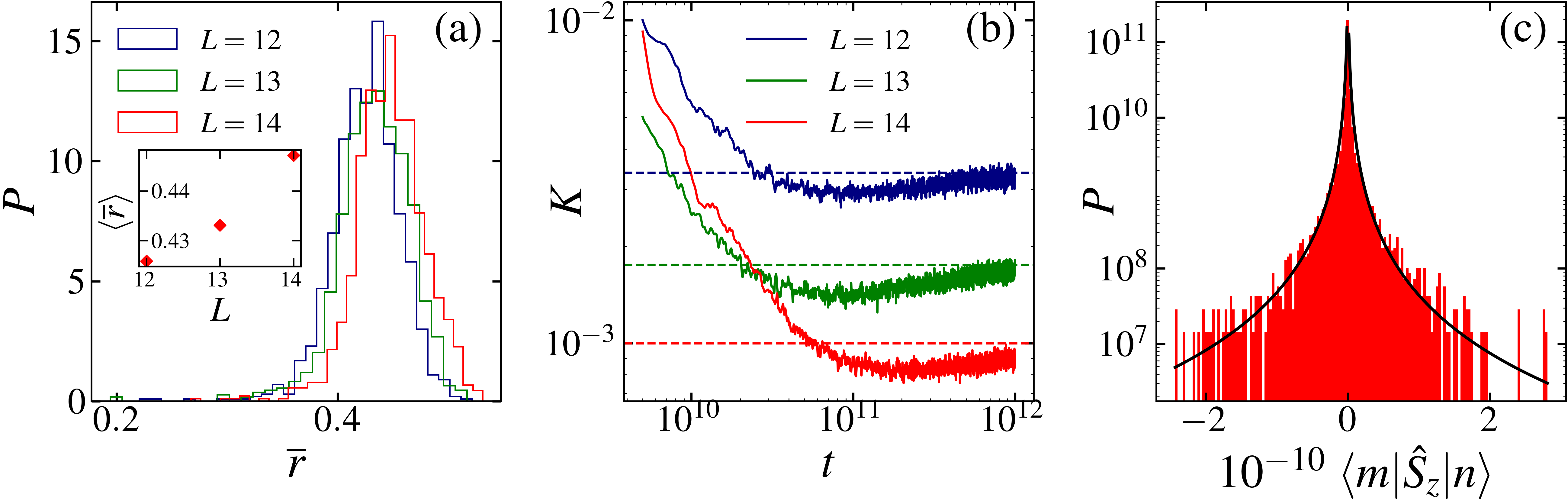}
    \caption{Analysis of (a)-(b) spectral correlations and (c) off-diagonal ETH for the most populated energy band of the perturbed Hamiltonian in Eq.~(2) of the main text with $\hat{V}=h_i \sigma_i^x$ , random numbers $h_i \in [-\delta,\delta]$ and $\delta=10^{-4}$.
    (a) Distribution of the mean value of the ratio of consecutive level spacings, $\bar{r}$, for various disorder realizations and their average vs system size $L$ in the inset. 
    (b) Spectral form factor for different system sizes; dashed line indicates the saturation value.
    (c) Distribution of the off-diagonal elements of $\hat{S}_z$ for 200 levels in the middle of the energy band, $L=13$. The solid line represents a log-normal distribution.   
    }
    \label{fig03}
\end{figure}

\subsection{Class (ii): Extensive one-body}
This class of perturbations includes both homogeneous and inhomogeneous field terms in both directions, applied to all sites.

{\bf For homogeneous perturbations}, as in the form of $ \sum_{i=1}^L \hat\sigma_i^x$ or $ \sum_{i=1}^L \hat\sigma_i^z$, the perturbed Hamiltonian still retains the original $SU(2)$ symmetry by conserving the total spin $\hat{S}^2$. This leads to the organization of the spectrum into more degenerate bands, thereby trivially extending the integrability of the unperturbed system.   

{\bf For inhomogeneous perturbations} in the transverse direction, $\sum_{i=1}^L h_i  \hat\sigma_i^z$, with $h_i \in [-\delta,\delta]$ being random, we find that the degeneracies are lifted inside a band. Nevertheless, the level statistics analysis reveals a clear Poissonian signature, implying integrability. Although the original $SU(2)$ conservation is lost, the first-order corrections do not reveal spectral correlations, which suggests integrability for the exact energy levels as well. 

For random field perturbations in the longitudinal direction, $\sum_{i=1}^L h_i  \hat\sigma_i^x$, the situation is more subtle. The situation becomes more intricate with inhomogeneous fields. Transverse inhomogeneous fields, for instance $\sum_i h_i \hat{\sigma}^z_i$, break symmetry and lift degeneracies, yet for either weak ($\delta \ll J$) or strong ($\delta \sim {\cal O}(J)$) disorder,  
we observe Poissonian level statistics for $\lambda_{n_0} $, indicating preserved integrability. We speculate that such behaviour is possibly related to Richardson-Gaudin integrability, as we demonstrate later that to be the case for fully-connected $XXZ$ Hamiltonian, where a $\sum_{i=1}^L h_i \hat{\sigma}_i^z$ perturbation allows one to construct an extensive set of conserved charges.

However, longitudinal inhomogeneous fields, $\sum_i h_i \hat{\sigma}^x_i$, lead to level statistics for $\lambda_{n_0} $ intermediate regime between Poisson and Wigner-Dyson distributions, suggesting level repulsion in the presence of possible  (quasi-)symmetries.
To further assess level repulsion and ETH in this setting, we return to the exact eigenvalues of the Hamiltonian in Eq.~(2) of the main text with $\sum_i h_i \hat{\sigma}^x_i$. We consider very weak random perturbations ($\delta = 10^{-4}$) and analyze in Fig.~\ref{fig03} different indicators of quantum chaos for the most populated energy band. The mean ratio of consecutive level spacings, $\bar{r}$, exhibits strong fluctuations across individual disorder realizations, as indicated by the relatively broad distributions in Fig.~\ref{fig03}(a). However, the ensemble average, $\langle \bar{r} \rangle$, increases with system size $L$, as evidenced in Fig.~\ref{fig03}(a) by a systematic shift toward larger values and narrowing of the distributions as $L$ increases. The inset confirms this trend, showing that $\langle \bar{r} \rangle$ grows with $L$.

In Fig.~\ref{fig03}(b) we examine the spectral form factor, defined as~\cite{MehtaBook}
\begin{align}
    K(t) = \left| \frac{1}{\mathcal{D}} \sum_{n=1}^{\mathcal{D}} e^{iE_n t} \right|^2,
\end{align}
where $\mathcal{D}$ is the Hilbert space dimension of the most populated energy band. This quantity sensitively captures spectral correlations by developing a dip–ramp–plateau structure (correlation hole) even in the presence of symmetries~\cite{Santos2020,Cruz2020}. The correlation hole, corresponding to the time interval with values of $K(t)$ below its saturation (dashed) line, is seen in Fig.~\ref{fig03}(b) for all three system sizes considered, with no sign of reduction of the relative depth with $L$. 

Figures~\ref{fig03}(a)-(b) suggest that integrability is broken for arbitrarily weak random longitudinal fields, yet quantum chaos is not fully developed. This is further supported by Fig.~\ref{fig03}(c), where off-diagonal ETH is probed with the distribution of the matrix elements of $\hat{S}_z$, as in Fig.~\ref{fig02}(g). Instead of the Gaussian form expected in many-body quantum chaos, we observe a log-normal distribution, indicative of structured eigenstates potentially shaped by residual (quasi-)symmetries.  These results point to the random longitudinal field perturbation as a subtle and nontrivial case deserving further theoretical and numerical exploration.

\subsection{Class (iii): Extensive two-body}
In this class of perturbations, we can consider nearest-neighbor interactions and beyond, as when incorporating the perturbation in $\alpha$ by changing it infinitesimally from $0$. The case $\alpha \rightarrow 0$ is shown in the main text. Let us consider here the case when the perturbation is nearest-neighbor and given by 
\[
\hat{V}^{\text{nn}} = \delta \sum_{k=1}^{L-1} \hat{\sigma}_k^x \hat{\sigma}_{k+1}^x.
\]

\begin{figure}[h]
    \centering
    \includegraphics[width=0.9\linewidth]{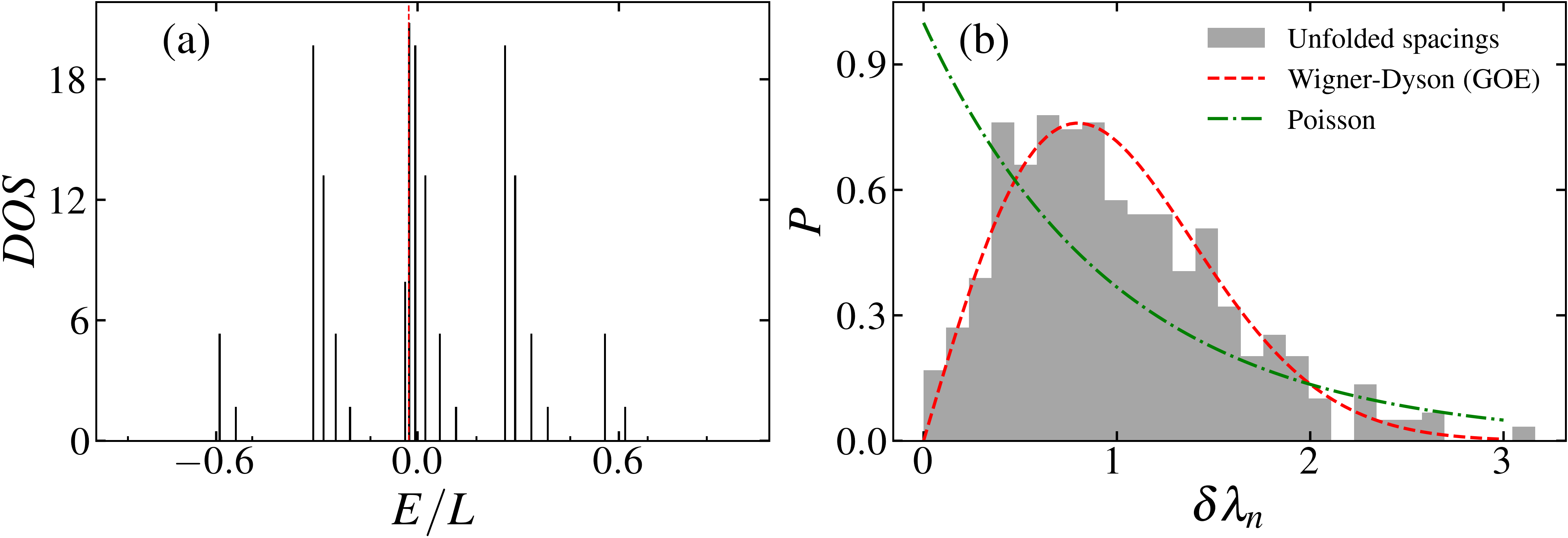}
    \caption{For the perturbation $\delta \sum_{k=1}^{L-1} \hat{\sigma}_k^x \hat{\sigma}_{k+1}^x$ with $\delta = 10^{-4}$, in panel (a) we show the density of states with the red-dashed line indicating one of the most populated bands for  $L=14$. In panel (b), the distribution of consecutive spacings for the first-order corrections in the band is shown to follow Wigner-Dyson statistics, underpinning the mechanism behind the onset of chaos at infinitesimal perturbation strength.  }
    \label{figSM:classIII-level-repulsion}
\end{figure}

We first examine the level statistics in the central one-third of the spectrum, resolved by both parity and inversion symmetries. Even for very small values of $ \delta $, we observe Wigner–Dyson statistics, indicative of quantum chaos. 
To further investigate the mechanism underlying this chaotic behavior, we construct the matrix $ \langle m_0 | \hat{V}^{\text{nn}} | n_0 \rangle $ within the most populated degenerate band of $ \hat{H}^{\alpha=0} $, restricted to the relevant symmetry sector. Diagonalizing this matrix, we obtain the first-order energy corrections $ \{ \lambda_{n_0}^i \} $. All the eigenvalues are found to be distinct, indicating that the degeneracy is completely lifted at first order in perturbation theory.
Even the set of first-order corrections $ \{ \lambda_{n_0}^i \} $ exhibits level repulsion. The distribution of the unfolded consecutive level spacings $ \delta \lambda_n^i = \lambda_{{n+1}_0}^i - \lambda_{n_0}^i $, after sorting the eigenvalues energetically, follows the Wigner-Dyson distribution. The emergence of level repulsion (see Fig.~\ref{figSM:classIII-level-repulsion}) already at the level of the first-order corrections provides a clear mechanism for the onset of chaos in the perturbed system. We note that the same holds for all perturbations in this class.
         
\section{Schur-Weyl Duality and coloring graphs for class(i) and class (ii) perturbations}
\subsection{Why is the spectrum banded?}
From Schur--Weyl duality, the Hilbert space
\begin{align}
\mathcal{H} = \mathbb{C}^d \otimes \mathbb{C}^d \otimes \cdots \otimes \mathbb{C}^d 
= \bigotimes_{i=1}^{L} \mathbb{C}^d
\end{align}
can be decomposed as 
\begin{align}
\mathcal{H} = \bigoplus_{\lambda} \left(\mathcal{V}_{\lambda}\otimes \mathcal{W}_{\lambda}\right),
\end{align}
where $\lambda$ labels Young diagrams, $\mathcal{V}_{\lambda}$ carries an irreducible representation of $SU(d)$, and $\mathcal{W}_{\lambda}$ carries the corresponding irreducible representation of the symmetric group $S_L$.

Let us denote the lattice of $L$ sites with local Hilbert space $\mathbb{C}^d$ as $\Lambda^d$. For any Hamiltonian satisfying
\begin{align}
P_{ij}\hat{H}P_{ij}=\hat{H}, \qquad \forall i,j\in\Lambda^d,
\end{align}
as defined in Eq.~(4) of the main text, the Hamiltonian is invariant under all pairwise permutations of the lattice sites, and the definition of the permutation operators is,
\begin{align}
    P_{ij} = \frac{1}{d}\mathbb{I}_d \otimes \mathbb{I}_{d} + \frac{1}{2} \sum_{\alpha=1}^{d^2 -1} T_i^{\alpha} \otimes T_j^{\alpha}, 
\end{align}
where $\{T^\alpha\} $s are the generators of $SU(d)$ group with $Tr[T^{\alpha} T^{\beta}]= 2 \delta_{\alpha\beta}$.
Since the transpositions $P_{ij}$ generate the symmetric group $S_L$~\cite{James2006representation}, the symmetry condition implies that $[\hat{H},\pi]=0$ for all $\pi\in S_L$. Consequently, by Schur's lemma, the Hamiltonian can be written in the block form
\begin{align} \label{eq:Schur's-lemma-block}
\hat{H}=\bigoplus_{\lambda}\left(\hat{h}_{\lambda}\otimes \mathbb{I}_{d_{\lambda}}\right),
\end{align}
where $\hat{h}_{\lambda}$ is the reduced Hamiltonian acting on the $SU(d)$ irreducible representation space $\mathcal{V}_{\lambda}$, and $\mathbb{I}_{d_{\lambda}}$ is the identity operator acting on the multiplicity space $\mathcal{W}_{\lambda}$ with $d_{\lambda}=\dim(\mathcal{W}_{\lambda})$.

This decomposition implies that the energy spectrum of $\hat{H}$ is obtained from the eigenvalues of each $\hat{h}_{\lambda}$, with every eigenvalue in a given sector $\lambda$ appearing with a degeneracy equal to $d_{\lambda}$, the dimension of the associated irreducible representation of the symmetric group $S_L$. Such a structure naturally leads to a banded energy spectrum, particularly when $\hat{H}$ can be expressed in terms of the Casimir operators or collective generators of $SU(d)$, since the energy scales are then determined by the global quantum numbers characterizing each representation sector, as in the case of the Hamiltonian in Eq.~(2) of the main text.

\subsection{Mental picture of all-to-all connected graph and coloring nodes: perturbations of class (i) and class (ii)}
It is useful to think of any Hamiltonian with the symmetry in Eq.~(4) of the main text as a fully connected graph of $L$ nodes with uniform color. Now let us consider one impurity is added at the lattice site $j_0 \in \Lambda^d$ such that the new Hamiltonian $\hat{\tilde{H}}$ would possess the symmetry structure
\begin{align}
    P_{ij} \hat{\tilde{H}} P_{ij} = \hat{\tilde{H}},~~\forall i,j \in \Lambda^d\setminus{j_0},
\end{align}
indicating $\hat{\tilde{H}}$ will now commute with every element from the symmetric permutation group of $L-1$ objects, denoted as $S_{L-1}\subset S_{L}$. Using Schur's lemma, we get the following block decomposition 
\begin{align}
    \hat{\tilde{H}} = \bigoplus_{\mu} \tilde{h}_{\mu} \otimes \mathbb{I}_{d_\mu},
\end{align}
where $\tilde{h}_{\mu}$ corresponds to the reduced Hamiltonian in partially broken $SU(d)$ space, and $\mathbb{I}_{d_\mu}$ acts on the multiplicity space $\mathcal{W}_{\mu}$ of dimension $d_{\mu}$. Here, $\mu$ denotes the irreducible representations of the subgroup $S_{L-1}$. This would be synonymus to coloring a single node in the fully connected graph of $\Lambda^d$. Now, by the same argument, one can think of adding $m$ number of impurities or $m$-body operator that reduces the symmetry structure to 
\begin{align}
    P_{ij} \hat{\tilde{H}} P_{ij} = \hat{\tilde{H}},~~\forall i,j \in \Lambda^d\setminus\{j_0,j_1,
    \ldots,j_m\},
\end{align}
which can still be block-decomposed using Schur's lemma since the Hamiltonian still commutes with elements from $S_{L-m}$. This automatically leads to the fact that if $m<<L$, the exponential degeneracy still prevails in the modified Hamiltonian, and we get a slightly lesser number of bands. This covers the perturbations of class (i), as described in the main text. 

If the perturbations are one-body field terms as in class (ii), then, depending upon whether they are random or uniform, we get two types of degeneracy preserving structures. If they are all homogeneous and $m=fL$ with $0\le f \le 1$, then we have two sets of permutation charges, and the fully-connected graph becomes a union of two graphs of distinct color. Instead, if the perturbations are random with strictly $f<1$, then the block-decomposition still survives since the Hamiltonian would still commute with $S_{(1-f)L}$, which hosts an extensive number of elements. A schematic representation is demonstrated in Fig.~\ref{fig:symmetry_breaking_graphs}.

The above structure can be understood in terms of representation-theoretic branching rules~\cite{James2006representation}. 
When $m$ random impurities are introduced, the permutation symmetry is reduced from $S_L$ to the subgroup $S_{L-m}$. 
Under this symmetry reduction, the irreducible representations $W_\lambda$ of $S_L$ decompose into irreducible representations of $S_{L-m}$ according to the branching rule
\begin{align}
W_\lambda \downarrow S_{L-m} =\bigoplus_{\mu} N_{\lambda\mu} W_\mu ,
\end{align}
where $N_{\lambda\mu}$ are non-negative integers specifying the multiplicities of the irreducible representations $W_\mu$ of $S_{L-m}$. 
Consequently, each spectral band associated with $\lambda$ splits into several subbands labeled by $\mu$. 
However, when $m \ll L$, the subgroup $S_{L-m}$ remains exponentially large, so the multiplicity spaces continue to host large degeneracies, resulting in a slightly fragmented but still highly banded spectrum.

\begin{figure}[ht!]
    \centering
    \begin{tikzpicture}[
        node distance=3.5cm,
        font=\sffamily\small,
        dot/.style={circle, fill=#1, inner sep=2.5pt},
        dot/.default=black,
        link/.style={gray!60, semithick} 
    ]

    \newcommand{\alltoall}[1]{
        \foreach \i in {1,...,8} {
            \coordinate (n\i) at ({(\i-1)*360/8}:1.2);
        }
        \foreach \i in {1,...,8} {
            \foreach \j in {\i,...,8} {
                \draw[link] (n\i) -- (n\j);
            }
        }
        #1
    }

    \node (center) at (0,0) {
        \begin{tikzpicture}
            \alltoall{
                \foreach \i in {1,...,8} \node[dot=black] at (n\i) {};
            }
        \end{tikzpicture}
    };
    \node[above=1.2cm of center, font=\large\bfseries]{};

    \draw[-{Stealth[scale=1.5]}, very thick, gray!80] (center.south) -- ++(-6,-3) coordinate (leftpos);
    \draw[-{Stealth[scale=1.5]}, very thick, gray!80] (center.south) -- ++(0,-3) coordinate (midpos);
    \draw[-{Stealth[scale=1.5]}, very thick, gray!80] (center.south) -- ++(6,-3) coordinate (rightpos);

    \begin{scope}[shift={(leftpos)}, yshift=-1cm]
        \alltoall{
            \node[dot=blue!80!black] at (n1) {};
            \node[dot=blue!80!black] at (n2) {};
            \foreach \i in {3,...,8} \node[dot=black] at (n\i) {};
        }
        \node[below=1.5cm, align=center, font=\bfseries] {Class (i): $m \ll L$ \\ \normalfont (Few Impurities)};
    \end{scope}

    \begin{scope}[shift={(midpos)}, yshift=-1cm]
        \alltoall{
            \foreach \i in {1,...,4} \node[dot=red!70!black] at (n\i) {};
            \foreach \i in {5,...,8} \node[dot=black] at (n\i) {};
        }
        \node[below=1.5cm, align=center, font=\bfseries] {Class (ii): Homogeneous \\ \normalfont ($S_{L/2} \times S_{L/2}$)};
    \end{scope}

    \begin{scope}[shift={(rightpos)}, yshift=-1cm]
        \alltoall{
            \node[dot=yellow!80!orange] at (n1) {};
            \node[dot=orange!80!red] at (n2) {};
            \node[dot=red!80!black] at (n3) {};
            \node[dot=purple!80!black] at (n4) {};
            \foreach \i in {5,...,8} \node[dot=black] at (n\i) {};
        }
        \node[below=1.5cm, align=center, font=\bfseries] {Class (ii): Partially Random \\ \normalfont (Residual $S_{(1-f)L}$)};
    \end{scope}

    \end{tikzpicture}
    \caption{Schematic representation of symmetry breaking in a fully connected graph of $L$ nodes. The central graph represents the pure $S_L$ symmetric Hamiltonian. The branches illustrate the reduction to $S_{L-m}$, $S_{L/2} \times S_{L/2}$, and $S_{(1-f)L}$ subgroups due to the introduction of various impurity classes.}
    \label{fig:symmetry_breaking_graphs}
\end{figure}

\section{Another example with long-ranged XXZ chain and connection to Richardson-Gaudin integrability for disordered fields}

We consider the following long-ranged XXZ  Hamiltonian 
\begin{align}\label{eq:XXZ-Hamiltonian}
    \hat{H}_{\mathrm{XXZ}} = \mathcal{N}_\alpha \sum_{i>j} \frac{1}{|i-j|^\alpha} \left( J\hat\sigma_i^x\hat\sigma_{i+1}^x + J\hat\sigma_i^y\hat\sigma_{i+1}^y + \Delta J\hat\sigma_i^z\hat\sigma_{i+1}^z    \right), 
\end{align}
where for any value of the long-ranged exponent $\alpha$, the $U(1)$ conservation $[\frac{1}{L}\sum_{i=1}^L\hat\sigma_i^z,\hat{H}_{\mathrm{XXZ}} ]=0$ is there. In the integrable limit, $\alpha=0$, the Hamiltonian $\hat{H}_{\mathrm{XXZ}}^{\alpha=0}$ permits the total spin conservation as well. Moreover, the Hamiltonian $\hat{H}_{\mathrm{XXZ}}^{\alpha=0}$ falls under the parent symmetry class in Eq.~(4) of the main text. 

\begin{figure}
    \centering
    \includegraphics[width=1\linewidth]{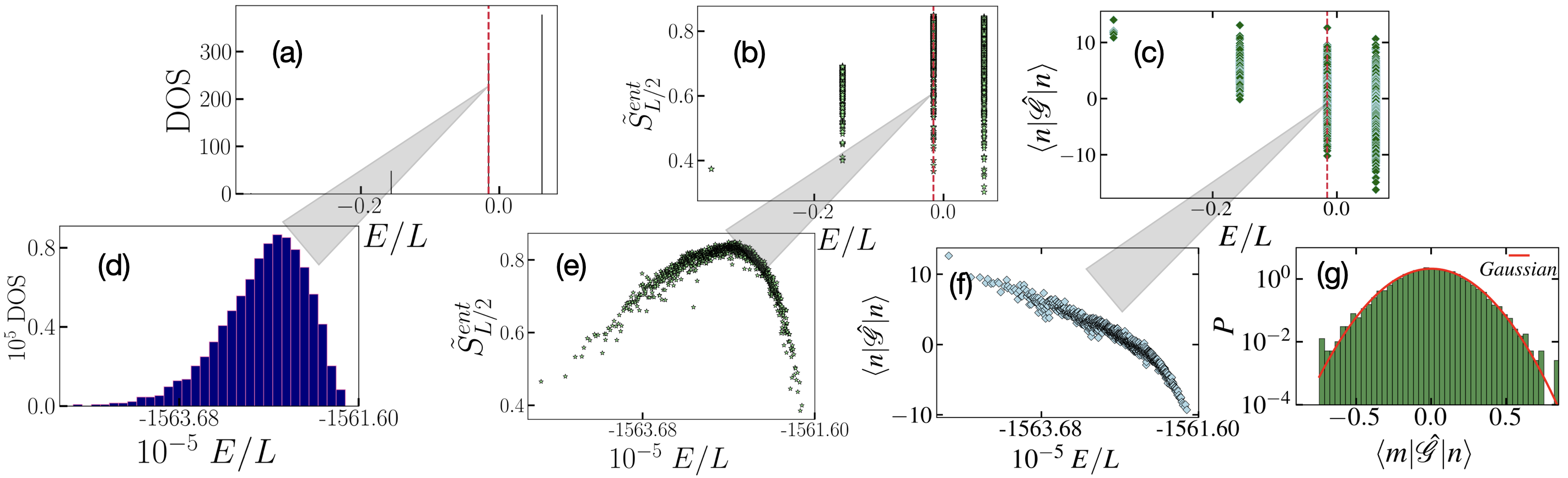}
    \caption{We report a similar finding of fragmented ETH as in Fig.~2 of the main text, but for the Hamiltonian considered in Eq.~\eqref{eq:XXZ-Hamiltonian} with $ L=16$, $\alpha=10^{-4},J =-1,~ \Delta=0.5$ in the $\hat{S}^z =0$ sector. The top panels show results for the full energy spectrum, with the vertical red line marking the most populated band analyzed in the bottom panels. (a) and (d): Density of states, (b) and (e): Entanglement entropy, (c) and (f): Eigenstate expectation values of $hat{\mathcal{G}} = \sum_{i=1}^{L-1} (\hat{\sigma}_i^x \hat{\sigma}_{i+1}^x  + \hat{\sigma}_i^y \hat{\sigma}_{i+1}^y)$, and (g): Off-diagonal elements of $\hat{\mathcal{G}}$ for 200 eigenstates in the middle of the selected energy band. (f) and (g): Diagonal and off-diagonal ETH, respectively, are satisfied. Parity and inversion symmetries are taken into account.}
    \label{fig:XXZ-feth}
\end{figure}

Similar to the observations in FIG.~2. of the main text, we find a qualitatively similar behavior for the XXZ Hamiltonian in terms of the emergence of ETH in individual bands when $\alpha$ is tuned from $0$ to a representative small value $10^{-4}$. To demonstrate this clearly, we consider the operator $\hat{\mathcal{G}} = \sum_{i=1}^{L-1} (\hat{\sigma}_i^x \hat{\sigma}_{i+1}^x  + \hat{\sigma}_i^y \hat{\sigma}_{i+1}^y)$ and perform the standard ETH tests of the main text and report in Fig.~\ref{fig:XXZ-feth}. 

\begin{figure}[ht!]
    \centering
    \includegraphics[width=0.5\linewidth]{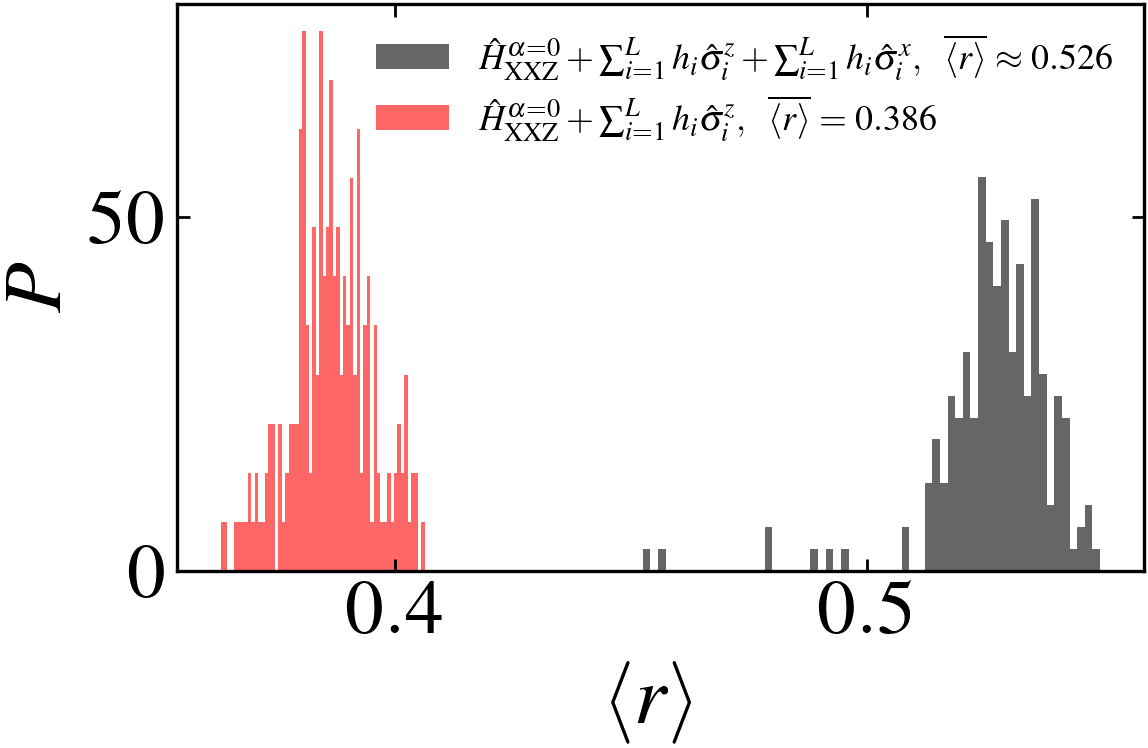}
    \caption{We show the distribution of the level-spacing ratio  $\langle r \rangle$ defined in the main text, over $200$ disorder realizations for $\hat{H}_{\mathrm{XXZ}}^{\alpha=0} + h_i \hat{\sigma}_i^z$ (red) and $\hat{H}_{\mathrm{XXZ}}^{\alpha=0} + h_i \hat{\sigma}_i^x + h_i \hat{\sigma}_i^z$ (black). In the first case, the system size considered is $L=14$ within the $\hat{S}_z=0$ sector, and the spectrum exhibits Poisson statistics across all realizations. In contrast, for the latter case the system size is $L=12$, and the spectrum displays Wigner–Dyson statistics for almost all realizations. The parameters used are $J=-1$, $\Delta=0.5$, and $\delta=1$.}
    \label{fig:XXZ-disordered-class-II}
\end{figure}

We note that the effect of class (i) and class (ii) perturbations (namely, the homogeneous case) is going to be identical even in this case due to the same structure of the permutation charges. However, we point out that for the disordered case of 1-body perturbations, the situation folds in a similar fashion to that of the one reported in the main text with the Hamiltonian in Eq.~(2). For a random perturbation of the form $h_i \hat{\sigma}_i^z$ with $h_i \in [-\delta,\delta]$ to $\hat{H}_{\mathrm{XXZ}}^{\alpha=0}$, the level spacing ratio as defined earlier shows Poissonian statistics across multiple disorder realizations. Moreover, only with the added perturbation in another direction of the form $h_i \hat{\sigma}_i^x$, the system $ \hat{H}_{\mathrm{XXZ}}^{\alpha=0} + h_i \hat{\sigma}_i^x+ h_i \hat{\sigma}_i^z$ shows signatures of Wigner Dyson statistics. We demonstrate this for the representative disorder strength $\delta=1$, see Fig.~\ref{fig:XXZ-disordered-class-II}. 

Moreover, the reason behind the Poisson statistics for a single directional field $(i.e.,~ \sum_{i} h_i \hat\sigma_i^{x,y,z})$ is that with a unidirectional disordered field, the system belongs to the Richardson-Gaudin class of integrable systems. The original Richardson model is of the following form 
\begin{align}
    \hat{H}_{RG} & = \sum_{i} h_i \hat{S}_i^z + g \sum_{ i,j=1}^L \hat{S}_i^+\hat{S}_j^- \\
    & = \sum_{i} h_i \hat{S}_i^z + g \sum_{ i,j=1}^L(\hat{S}_i^x \hat{S}_j^x + \hat{S}_i^y \hat{S}_j^y),
\end{align} 
which supports an extensive number of charges $\{\hat{R}_i: [\hat{H}_{RG},\hat{R}_i ] =0, \forall i=1,\ldots,L\}$ (see \cite{Cambiaggio1997,Dukelsky2001}) of the form 
\begin{align}
    \hat{R}_i = \hat{S}_i^z + g \sum_{j\neq i} ^L \frac{1}{h_i -h_j } \left[ \frac{1}{2}(\hat{S}_i^+\hat{S}_j^- + \hat{S}_i^-\hat{S}_j^+) + \hat{S}_i^z \hat{S}_j^z \right]. 
\end{align}
Now, let us note the following commutation relation
\begin{align}
    \left[\sum_k \hat{S}_k^z , R_i\right] & = J \sum_{j\neq i} ^L \frac{1}{2(h_i -h_j) } \left\{ \underbrace{\left[\sum_k \hat{S}_k^z, \hat{S}_i^+\hat{S}_j^-  \right]}_{=0} +  \underbrace{\left[\sum_k \hat{S}_k^z, \hat{S}_i^-\hat{S}_j^+  \right]}_{=0} \right\}\\
    & = 0,
\end{align}
which trivially leads to the condition that for any function of the collective operator $\sum_k \hat{S}_k^z$, we have 
\begin{align}
    [f(\sum_k \hat{S}_k^z), R_i]=0.
\end{align}
This fact indicates that any deformation of the form $f(\sum_k \hat{S}_k^z)$ to the Hamiltonian $\hat{H}_{RG}$ will also have the same set of conserved charges, since 
\begin{align}
    \left[\hat{H}_{RG} + f(\sum_k \hat{S}_k^z), R_i\right] =0 .
\end{align}
For the particular choice $f(\sum_k \hat{S}_k^z) = \frac{\Delta}{L}(\sum_k \hat{S}_k^z)^2 $, we have 
\begin{align}
    \hat{H}_{RG}(g=J/4L) +\frac{\Delta}{4L}(\sum_k \hat{S}_k^z)^2 = \hat{H}_{\mathrm{XXZ}}^{\alpha=0} + \sum_{i=1}^L h_i \hat{S}_i^z .
\end{align}
Therefore, $\hat{H}_{\mathrm{XXZ}}^{\alpha=0} + \sum_{i=1}^L h_i \hat{S}_i^z$ is clearly an integrable Hamiltonian belonging to the Richardson-Gaudin family. This explains the observed Poisson statistics. Once the field in the other direction has been added as well, the system moves away from integrability and Wigner-Dyson statistics emerges.

\phantomsection
\label{LastBibItem}
\end{document}